\documentclass{ws-p8-50x6-00}

\begin{document}

\title{Status of Nucleon Resonances with Masses $M<m_N+m_{\pi}$}

\author{R. Beck$^1$, S.N. Cherepnya$^2$, L.V. Fil'kov$^2$,
V.L. Kashevarov$^2$,\\ M. Rost$^1$, and Th. Walcher$^1$}

\address{$^1$ Institut f\"{u}r Kernphysik, Johannes Gutenberg-Universit\"{a}t, 
Mainz, Germany}

\address{$^2$ Lebedev Physical Institute, Leninsky Prospect 53, 117924 Moscow,
Russia\\E-mail: filkov@x4u.lpi.ruhep.ru}

\maketitle

\abstracts{
We discuss different interpretations of the peaks observed a few years ago by
Tatischeff {\it et al.} \cite{tat} in the missing mass spectra of the reaction 
$pp\to \pi^+ pX$, which were declared 
as new exited nucleon states with small masses. A study of the
possible production of such states in the process $\gamma p\to \pi^+N^*
\to \pi^++\gamma\gamma n$ by analysing the invariant mass spectrum of 
$\gamma\gamma n$ is proposed. It is shown that the experimental data obtained
at MAMI-B allows to analyse this process and to get information about
an existence of exited nucleon states with small masses.}

A few years ago
three narrow bumps have been observed in missing mass spectra of the
reaction $pp\to p\pi^+ X$ by Tatischeff {\it et al.} \cite{tat} at 
$M_X=1004$, 1044, and 1094 MeV. These bumps were interpreted as new 
nucleon resonances $N^*$. The values of masses $M_X=1004$ and 1044 MeV 
are below $m_N+m_{\pi}$ and so these states can decay with an emission 
of photons only. If they decay into $\gamma N$ then these resonances 
have to contribute to the Compton scattering on nucleon. However, the 
analysis \cite{lvov} of the existing experimental data on this processes 
completely excluded such $N^*$ as intermediate states in the Compton 
scattering on the nucleon.

In Ref. \cite{kob} it was assumed that these states belong to the totally
antisymmetric \underline{20}-plet of the spin-flavor SU(6)$_{FS}$.
Such a $N^*$ can transit into nucleon only if two quarks from $N^*$ 
participate in the interaction. Then the simplest decay of $N^*$ with the
masses 1004 and 1044 MeV is $N^*\to \gamma\gamma N$. This assumption could
be checked, in particular, by investigating the reaction 
$\gamma p\to\gamma X$ or $\gamma p\to\pi X$ in the photon energy region 
about 800 MeV.

Another interpretation of the states found in work \cite{tat} was 
suggested in Refs.~\cite{fil1,fil2}. In these works the reaction 
$pd\to p+pX_1$ has been studied with the aim of searching for
supernarrow dibaryons (SND), decay of which into two nucleon is forbidden by 
the Pauli exclusion principle. Three peaks have been observed in invariant 
mass spectra of $pX_1$ states at $M_{pX_1}=1904\pm 2$, $1926\pm 2$, and
$1942\pm 2$ MeV (see Fig.1a). 
\begin{figure}[t]
\epsfxsize=19pc
\epsfbox{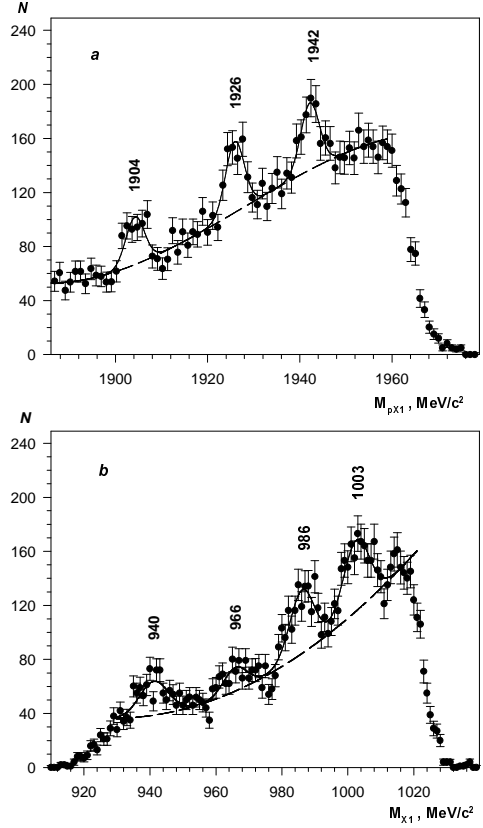}
\caption{The invariant mass $M_{pX_1}$ (a) and missing mass $M_{X_1}$ (b) spectra}
\end{figure}
The analysis of the angular distribution of
the protons from the decay of $pX_1$ states showed that the peaks found can
be explained as manifestation of the SNDs. Additional information about the 
nature of these states have been obtained by analysing the missing mass
$M_{X_1}$ spectra. If the observed state is a dibaryon decaying mainly into
two nucleons then $X_1$ is a neutron and $M_{X_1}$ has to be equal to the 
neutron mass $m_n$. If the value of $M_{X_1}$, obtained from experiment,
differs essentially from $m_n$ then $X_1=\gamma+n$ and the found state is
SND.

Fig.1b demonstrates the missing mass $M_{X_1}$ spectrum obtained in 
Refs~\cite{fil1,fil2}. As is seen from this figure, besides the peak at the neutron 
mass, which caused by the process $pd\to p+pn$, the resonancelike behavior 
of the spectrum is observed at $966\pm 2$, $986\pm 2$, and $1003\pm 2$ MeV.
These values of $M_{X_1}$ coincide with the simulated ones and differ 
essentially from the neutron mass. Hence, the dibaryons found are really
SNDs.
 
It should be noted that the peak at $M_{X_1}=1003\pm 2$ MeV corresponds to
the bump observed in Ref~\cite{tat}. Taking into account the found connection 
between SNDs and resonancelike states $X_1$, authors of the 
works~\cite{fil1,fil2} assumed
that the peaks from Ref~\cite{tat} at 1004 and 1044 MeV are not the exited 
nucleons, but they are the resonancelike states $X_1=\gamma+n$ caused by
possible existence and decay of the SNDs with the masses 1942 and 1982 MeV.
Such $X_1$ are not real resonances and cannot give contribution to the Compton
scattering on the nucleon.

However, a SND can also decay into $NN^*$. Unfortunately, the 
experiment~\cite{fil1,fil2} could not unambiguously discriminate an exited 
nucleon from $X_1$ state. To clarify this question we propose to study the 
exited nucleon production in the process
\begin{equation}
\gamma+p\to \pi^+ +N^*\to \pi^+ +\gamma\gamma n
\end{equation}
by analysing the invariant mass spectrum of the $\gamma\gamma n$ at the 
incident photon energy from 537 up to 817 MeV. The data on this process
can be obtained from the experiment on the radiative $\pi^+$ meson 
photoproduction from the proton, which has been carried out at MAMI-B.
In this experiment $\gamma$, $\pi^+$, and $n$ were detected, so we have 
enough data to reconstruct the invariant mass of the $\gamma\gamma n$
from 985 up to 1075 MeV. The lowest value of $M_{\gamma\gamma n}$ is due 
to the threshold of the final photon energy (15 MeV) and the highest one
corresponds to the sum of masses of the neutron and $\pi^0$ meson.

The main background processes are $\gamma p\to\pi^0\pi^+n$ and
$\gamma p\to\pi^0\pi^0\pi^+n$. To suppress the contribution of the double 
pion photoproduction, we will consider the invariant mass of two photons
$M_{\gamma\gamma}<110$ MeV.
As the $\pi^+$ meson and $N^*$ must fly in the
same plain, we have an additional condition on the difference of pion and
$N^*$ azimuthal angles: 
$160^{\circ}<|\phi_{\pi^+}-\phi_{N^*}|<200^{\circ}$. 
This condition allows to suppress the contribution of the triple pion 
photoproduction.

The results of the simulation of the $N^*$ production in the process (1),
at the conditions of the radiative $\pi^+$ meson photoproduction from 
proton experiment and for an exposition time of 100 hours, are shown in 
Fig.2. 
\begin{figure}[t]
\epsfxsize=16pc
\epsfbox{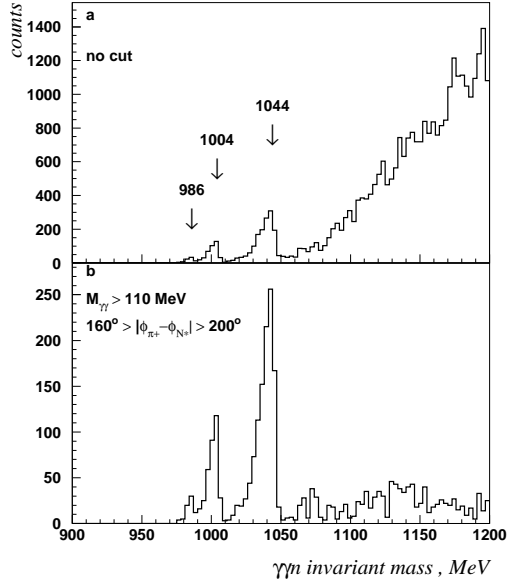}
\caption{ Invariant mass spectra of $\gamma\gamma n$. (a) --  without any cuts;
(b) -- taking into account conditions $M_{\gamma\gamma}<110$ MeV and
$160^{\circ}<|\phi_{\pi^+}-\phi_{N^*}|<200^{\circ}$.}
\end{figure}  
The calculations obtained without any cuts are presented in Fig.2a.
Fig.2b demonstrates the final result after both cuts.
As is seen from this figure, the background can be well suppressed and,
if $N^*$ states exist, they would be well recognizable.

In the experiment on radiative $\pi^+$ meson photoproduction we had about
1000 hours of exposition time. As result, we expect to get about 600, 
3500, and 12000 events for the exited nucleon states with the masses 986, 
1004, and 1044 MeV, respectively. So this experiment can give important
information about possibility of existence of exited nucleon states with 
small masses.

This work was supported in part by RFBR, grant No 00-02-04014 NNIO\_a,   
and Deutsche Forschungsmeinschaft (SFB 443).

\end{document}